\documentclass[10pt,conference]{IEEEtran}

\usepackage{cite}
\usepackage{amsmath,amssymb,amsfonts}
\usepackage{algorithmic}
\usepackage{graphicx}
\usepackage{textcomp}
\usepackage{xcolor}

\usepackage{multirow}
\usepackage{makecell}
\usepackage{ragged2e}
\usepackage{booktabs}
\usepackage{microtype}
\usepackage{tabularx}
\newcolumntype{R}{>{\raggedright\arraybackslash}X}
\usepackage[utf8]{inputenc}
\usepackage{placeins}
\usepackage[english]{babel}
\usepackage{csquotes}

\usepackage{amsthm}
\newtheorem{hyp}{Hypothesis}

\def\BibTeX{{\rm B\kern-.05em{\sc i\kern-.025em b}\kern-.08em
    T\kern-.1667em\lower.7ex\hbox{E}\kern-.125emX}}
\begin{document}

\title{To get good student ratings should you only teach programming courses? Investigation and implications of student evaluations of teaching in a software engineering context}

\author{\IEEEauthorblockN{Antti Knutas}
\IEEEauthorblockA{\textit{LUT University}\\
Lappeenranta, Finland \\
antti.knutas@lut.fi}
\and
\IEEEauthorblockN{Timo Hynninen}
\IEEEauthorblockA{\textit{South-Eastern Finland University of Applied Sciences}\\
Mikkeli, Finland \\
timo.hynninen@xamk.fi}
\and
\IEEEauthorblockN{Maija Hujala}
\IEEEauthorblockA{\textit{LUT University}\\
Lappeenranta, Finland \\
maija.hujala@lut.fi}}

\maketitle

\begin{abstract}
Student evaluations of teaching (SET) are commonly used in universities for assessing teaching quality. However, previous literature shows that in software engineering students tend to rate certain topics higher than others: In particular students tend to value programming and software construction over software design, software engineering models and methods, or soft skills. We hypothesize that these biases also play a role in SET responses collected from students. The objective of this study is to investigate how the topic of a software engineering course affects the SET metrics. We accomplish this by performing multilevel regression analysis on SET data collected in a software engineering programme. We analyzed a total of 1295 student evaluations from 46 university courses in a Finnish university. The results of the analysis verifies that the student course evaluations exhibit similar biases as distinguished by previous software engineering education research. The type of the course can predict a higher SET rating. In our dataset, software construction and programming courses received higher SET ratings compared to courses on software engineering processes, models, and methods. 
\end{abstract}


\section{Introduction}
There are established recommendations of what should be included in a software engineering curriculum~\cite{ardis2015se} and professionals have established a rough, evolving consensus of what is included in the field of software engineering~\cite{bourque2014guide}. However, while the software engineering education community and practitioners might agree on the content of the curriculum, students completing these programs may not share this point of view. In fact, studies have shown that in software engineering and computing related fields students emphasize the importance of programming, especially at the start of their studies, and might devalue other parts of degree programs~\cite{ivins2006software, hewner_undergraduate_2013, hewner_how_2014, gold2019software}. 

Student perceptions of the usefulness of the course topics is important, not only for purposes of the student's professional growth and understanding of the field, but also for meaningful dialogue about the content of the study program. One of the most common ways universities engage in this dialogue is through their quality control processes, which often use student evaluations of teaching (SET) as the main source of data. It is common for universities to use student evaluations of teaching as an indicator of quality for both the teaching material and teachers themselves~\cite{shevlin2000validity, zabaleta2007use}. 

Sometimes universities connect teaching faculty performance evaluation directly to student evaluation of teaching~\cite{zabaleta2007use}. The evaluations can affect the career prospects of the teaching personnel because the administration can use the data collected from SET questionnaires for decisions such as tenure, promotion and merit-pay~\cite{spooren2013validity}. But; if students value certain courses and topics over others, is there a built-in bias to the SET metrics?

In addition to being treated as an objective data source by administrative departments, SET is also being used by teachers to reflect on their teaching practises~\cite{winchester2011exploring}. For both reasons, it is an important field of study. However, while SET has been researched in general, it has received little attention in the field of software engineering education (SEE). In this paper, we address the research gap by exploring the statistical connection between a software engineering course type and the SET.

We accomplish our research goal by performing multilevel regression analysis on 1295 course evaluations from 46 software engineering courses, gathered between autumn 2017 and spring 2020 in a Finnish university. Our main research question is as follows:

\textit{How does the type of course affect student evaluation of teaching in software engineering courses?}

The rest of this paper is structured as follows. Section II presents the related work on student evaluation of teaching, and discusses the state of the art of SET in software engineering education. Section II also establishes the research gap. Section III presents the research approach, detailing the hypotheses, data collection, and data analysis methods. The main results of the study are presented in Section IV, while Section V discusses these results. Finally, Section VI concludes the paper.

\section{Background}

\subsection{Student evaluation of teaching (SET)}

SET is a commonly used measure for teaching quality in higher education~\cite{marsh1987students, uttl2017meta, wallace2019state}. In fact, according to many articles, SET is the most common method to evaluate faculty’s teaching performance in higher education institutions~\cite{clayson2009student,hoel2019bother,kember2002does,spooren2010credibility,spoorena2012participates, wallace2019state}. It is surprising that SET is the only widely used method for assessing teaching quality as there exists many different methods of assessing teacher and teaching quality besides student evaluations, including peer-rating, self-evaluation, student interviews, learning outcome measures and teaching portfolios~\cite{berk2005survey}.

SET is also a controversial measure for teaching quality, as student ratings of teaching and student learning are not related~\cite{uttl2017meta}. Previous research has shown that SET is a multidimensional concept~\cite{marsh1980influence,marsh1984students,marsh1987students, marsh2001distinguishing, marsh2007students, marsh2009exploratory}, whose validity for formative or summative purposes remains questioned~\cite{spooren2010credibility, spooren2013validity}. There is ample evidence that various student, teacher and course characteristics play a role in SET~\cite{marsh1987students, pounder2007student, spooren2013validity, wachtel1998student}: For example, on average female students provide higher SET ratings than males~\cite{kohn2006role}. Some evidence also indicates that older students appear to provide higher SET ratings \cite{spooren2010credibility}. Teachers’ charisma appears to be associated strongly with perceived teaching ability~\cite{shevlin2000validity}, and physically attractive teachers are likely to receive higher SET ratings~\cite{gurung2007looking, hamermesh2005beauty, riniolo2006hot}. 

As for how teaching methods affect SET ratings, previous studies’ results are somewhat ambiguous. Some evidence indicates that students rate online courses lower than face-to-face courses~\cite{lowenthal2015student}, whereas results from Carle \cite{carle2009evaluating} indicate no differences between instruction methods except for teachers with racial minority status. A characteristic often found to be important is course rigor, which Clayson~\cite{clayson2009student} states is associated negatively with SET ratings in general. Rigor has been measured, e.g., through students’ perceptions of course difficulty~\cite{centra2003will, marsh2000effects, remedios2008liked, ting2000multilevel}, course workload~\cite{centra2003will, marsh2001distinguishing, marsh2000effects}, and course pace~\cite{centra2003will, marsh2000effects}. 

\subsection{SET in computer science and software engineering education} \label{sec:set_cs_sw}

Existing work on SET in a computer science education (CSE) or SEE is scarce~\cite{garcia2010evaluation}, and to our knowledge notable research efforts in the cross-section of SET and computing education have not been made in the past decade. 

There are some recent works that deal with SET in software engineering and computing education. Kavalchuk et al.~\cite{kavalchuk2020empirical} analyzed data from RateMyProfessor.com to distinguish the qualities of popular CS and SWE instructors. In a similar vein Carbone and Ceddia mined student evaluations for improvement areas in the ICT field~\cite{carbone2012common}. 

The concept of SET has been used implicitly in many computing education papers: In these works teaching tools, pedagogical interventions, or curricular implementations have been validated by using student feedback data. Often student evaluations are used by researchers in the CSE/SEE communities to validate the design of courses.  For example, among (the many) recent software engineering papers the study of Ralph~\cite{ralph2018re} evaluates the implementation of a course in Software Project Management, and Agneli et al.~\cite{angeli2020constructivist} a graduate course in web service design.

The low number of studies related to the use of SET in SEE is an essential research gap, since previous studies have shown that software engineering and computer science students value different areas of their fields differently. Research on student misconceptions show that students emphasize hands-on programming over other subfields, such as design or engineering processes~\cite{ivins2006software}.

Furthermore, existing research presents evidence that students consider particular skills as more central to software engineering or computing. For example, Ivins et al.~\cite{ivins2006software} found that writing computer programs was emphasized as a skill compared to requirements engineering or design. Similarly, Gold-Veerkamp found~\cite{gold2019software} that implementation is considered strongly part of software engineering, whereas some parts of design, requirements engineering, and quality assurance were not. Hewner~\cite{hewner_undergraduate_2013} had similar outcomes in a related field, computer science, where the role of programming was emphasized over computer science theory.

\section{Methods}
\subsection{Research approach and hypotheses}

In this study we examine whether student evaluations of software engineering courses vary between different course types. More specifically, we address a part of the research gap presented in Section \ref{sec:set_cs_sw} by investigating whether the student evaluations of teaching in software engineering courses reflect the fact that students tend to value certain course topics over others. 

We base our course type categorization on the Guide to the Software Engineering Body of Knowledge (SWEBOK) \cite{bourque2014guide}. SWEBOK was selected because the software engineering curriculum at the studied university follows the ACM/IEEE 2014 joint task force guidelines~\cite{ardis2015se}, which in turn have been based on empirical research and existing knowledge bases, such as SWEBOK~\cite{bourque2014guide}. Furthermore, several other analyses in the field apply SWEBOK~\cite{garousi_understanding_2020}.

Our hypotheses are as follows:
\begin{hyp}
The type of course (based on SWEBOK categorization) affects student evaluation of teaching in software engineering courses.
\end{hyp}
\begin{hyp}
Courses related to \textit{software construction and programming} provide higher SET ratings than courses related to other knowledge areas.
\end{hyp}

\subsection{Data} 
We test our hypotheses using student feedback data from the feedback surveys carried out at a Finnish university between academic years 2017-2018 and 2019-2020. The data was collected through two slightly different student feedback questionnaires\footnote{Survey questions are available at https://doi.org/10.5281/zenodo.4519256}: one for the academic year 2017-2018 and the other for 2018-2019 and 2019-2020.

The first questionnaire (2017-2018) comprised of five Likert-scale questions assessing students’ motivation, effort put into learning, workload, and teaching methods and course implementation in relation to perceived learning. Five open-ended questions were included as well.

The second questionnaire (2018-2019 and 2019-2020) comprised of four Likert-scale questions assessing students’ motivation, workload, and teaching methods and course as a whole in relation to perceived learning. In addition, four open-ended questions were included in the questionnaire. 

The survey questionnaires were sent to students via email after they completed the courses. The surveys mostly were sent to all students enrolled in the courses, but teachers can collect attendance and limit feedback surveys to only those students who attended classes. Responding was anonymous and voluntary for all students.

The sample is restricted to student feedback from software engineering courses with ten or more student feedback questionnaires filled out. The sample includes 415 responses from 16 courses in 2017-2018, 395 responses from 15 courses in 2018-2019, and 485 responses from 15 courses in 2019-2020. As four of the Likert-scale questions were identical, or very similar, between the two student feedback questionnaires used, we combined all responses from each three academic year into one data set. The combined data set consists of student feedback collected from 22 courses taught one to three times over the three academic years studied. The total number of course implementations is 46 and the total number student feedback questionnaires filled out is 1295.
\subsection{Measures}
\subsubsection{Dependent variable}
We carried out an exploratory factor analysis of the four SET items of the combined data set. Factors were extracted using principal factor analysis with promax rotation. A scree plot of eigenvalues was used to determine the optimal number of factors.

One factor was identified. Two items reflecting student's perceptions about the course and its teaching methods in relation to perceived learning had high loadings ($>$ 0.8) on this factor. The items exhibited good reliability (Cronbach's alpha $=$ 0.856), and they were averaged together to form a measure of student's \textit{learning experience} on a scale from 1 (the worse) to 5 (the best). This is our dependent variable in Hypotheses 1 and 2.

Two items - ‘My motivation in this course was (1 = very low; 5 = very high)’ and ‘The workload relative to the study credits awarded was (1 = very light; 5 = very heavy)’ - did not load on the factors and were used as single-item measures of student's \textit{motivation} and \textit{perceived workload}. These items serve as control variables in the analysis because the previous literature has found evidence that student's motivation and perceived workload play a role in SET. According to, for example, Griffin~\cite{griffin2004grading} and Wachtel~\cite{wachtel1998student}, students’ pre-course motivation or prior subject interest is positively associated with SET: interested students appear to give higher ratings. In turn, ``just right'' level of workload leads to better SET ratings~\cite{centra2003will,marsh2001distinguishing}.
\subsubsection{Course type}
\begin{table*}
\centering
  \caption{Course types and corresponding SWEBOK knowledge areas}
  \begin{tabular}{ll}
    \toprule
    \textbf{Course type}&\textbf{SWEBOK knowledge areas}\\
    \midrule
    A. SW construction and programming & SW construction\\
    & SW testing\\
    & SW maintenance \\
    \midrule
    B. SW engineering process, models and method & SW design\\
    & SW engineering models and methods\\
    & SW requirements\\
    & SW engineering process\\
    &SW quality\\
    \midrule
    C. Professional practices for SW engineering & SW engineering professional practice\\
    &SW engineering economics\\
    &SW engineering management\\
\bottomrule
\end{tabular}
\label{table:coursetype}
\end{table*}
We classified the 22 courses into three categories according to the course content. The categories are based on SWEBOK knowledge areas (see Table~\ref{table:coursetype}) and labeled as A) Software construction and programming, B) Software engineering process, models and methods, and C) Professional practices for software engineering. The categories are referred to here as \textit{course types}. Course type is used as an independent variable in the analysis.

\subsection{Analysis methods} 
We use multilevel regression analysis~\cite{raudenbush2002hierarchical} to address the research questions. The main reason for employing multilevel analysis is that the observations of the SET data presumably are not independent. Student evaluations are nested within course implementations and course implementations are nested within courses (see Figure 1). In other words, for example, SETs from the same course implementation presumably share more similarities than they do with SETs from other course implementations. Ignoring data clustering may lead to underestimated standard errors of regression coefficients and, thus, overly small p-values. Multilevel analysis takes into account this clustered structure of the data. In addition, it allows us to examine the relationships between variables at different levels of the data (course type and student's learning experience).
\begin{figure*}[ht]
   \centering
   \includegraphics[width=1\textwidth]{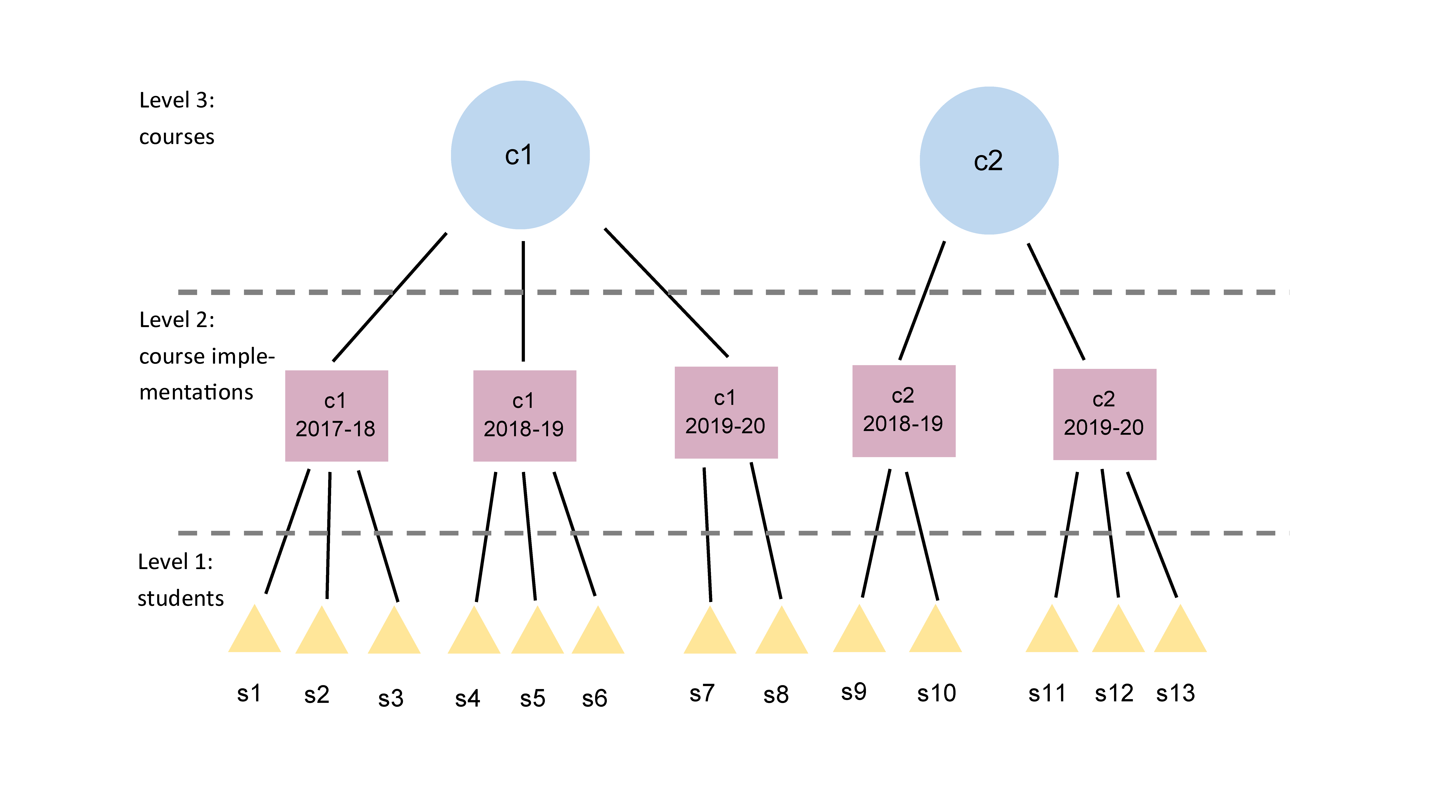}
   \caption{Illustration of the data structure}
   \label{fig:figure_levels}
\end{figure*}
 
Multilevel models allow for residual components at all levels – at course level (level 3 in Figure 1), course implementation level (level 2 in Figure 1) and student level (level 1 in Figure 1). However, in preliminary analyses we found out that the amount of level-3 variation is very small. Only 0.86 percent of the variance in the learning experience was situated at level 3 (course level). The rule of thumb is that if 5 percent or more of the variance is attributable to the level it should not be ignored~\cite{mehmetoglu2016applied}. Thus, we chose to ignore the third level and fit two level (student level and course implementation level) models instead.

Hypotheses 1 and 2 are jointly tested by fitting the following random coefficient model with student's learning experience (LE) as the outcome variable and student's motivation (MO), perceived workload (WL) and course type (COURSETYPE) as predictors: 

{\small
\begin{multline}
LE_{ij}=\beta_0+\beta_1MO_{ij}+\beta_2WL_{ij}+\beta_3WL^2_{ij} \\
+\beta_4COURSETYPE_j+u_{0j}+u_{1j}WL_{ij}+u_{2j}WL^2_{ij}+e_{ij} 
\end{multline}
}

The relationship between learning experience and perceived workload is assumed to be curvilinear as suggested, for example, by Centra~\cite{centra2003will}. In addition, the impact of perceived workload on the learning experience is assumed to differ between courses. 

We used Stata/SE 16.1 software for all analyses. 

\section{Findings}
\begin{table*}
\centering
  \caption{Descriptive statistics of the student level variables}
  \begin{tabular}{lccccc}
    \toprule
    \textbf{Variable}&\textbf{Mean}&\textbf{SD}&\textbf{Min}&\textbf{Max}&\textbf{n}\\
    \midrule
    Learning experience & 3.46 & 1.13 & 1 & 5 & 1282 \\
    Motivation & 3.67 & 1.06 & 1 & 5 & 1287 \\
    Perceived workload & 3.61 & 0.92 & 1 & 5 & 1280 \\
\bottomrule
\end{tabular}
\label{table:descriptive1}
\end{table*}
\begin{table*}
\centering
  \caption{Descriptive statistics of the course type}
  \begin{tabular}{lccc}
    \toprule
    \textbf{Course type}&\textbf{No. of course implementations}&\textbf{Percent}&\textbf{Mean learning experience}\\
    \midrule
    A. SW construction and programming & 27 & 58.70 & 3.63\\
    B. SW engineering process, models and method & 10 & 21.74 & 3.08\\
    C. Professional practices for SW engineering & 9 & 19.57 & 3.27\\
\bottomrule
\end{tabular}
\label{table:descriptive2}
\end{table*}

Descriptive statistics of the variables are presented in Tables~\ref{table:descriptive1} and~\ref{table:descriptive2}. As shown in Table~\ref{table:descriptive1}, the average learning experience (range $=$ 1-5, mean $=$ 3.46, SD $=$ 1.13) is slightly above the middle of the range indicating that, on average, the software engineering students have rather good learning experience. Furthermore, students' average motivation is 3.67 (range $=$ 1-5, SD $=$ 1.06) and the average perceived workload is 3.61 (range $=$ 1-5, SD $=$ 0.92). It thus seems that, on average the software engineering students are reasonably well motivated and, according to the them, the workload of the software engineering courses is not too high or too low.

Table~\ref{table:descriptive2} shows the proportion of course implementations within each course type and also the mean learning experience for each of them. As shown, the mean learning experience varies from 3.08 in SW engineering process, models and method courses to 3.63 in SW construction and programming courses. 

As a preliminary analysis, a one-way ANOVA was was conducted to compare the effect of course type on the learning experience. An analysis of variance showed that the effect was statistically significant, F(2,1279) $=$ 24.27, p $<$ 0.001 suggesting the need for a more thorough analysis of the role of course type in students' learning experience.

\begin{table*}
\centering
  \caption{Estimated parameters of the two-level random coefficient models predicting student's perceptions of learning
}
  \begin{tabular}{lcc}
    \toprule
     & \multicolumn{2}{c}{\textit{Learning experience}} \\ 
    \midrule
    Fixed effects & & \\
    \hspace{4pt} \textit{Intercept} & 3.591*** & (0.079) \\
        \hspace{4pt} \textit{Motivation} & 0.525*** & (0.025) \\
        \hspace{4pt} \textit{Workload} & -0.062 & (0.040) \\
        \hspace{4pt} \(Workload^2\) & -0.121*** & (0.028) \\
    \hspace{4pt} \textit{Course type} &  & \\
        \hspace{12pt} \textit{A} & & \\
        \hspace{12pt} \textit{B} & -0.353* & (0.149) \\
        \hspace{12pt} \textit{C} & -0.165 & (0.147) \\
    Random effects & & \\
        \hspace{4pt} \textit{var\_workload} & 0.021 & \\ 
        \hspace{4pt} \textit{var\_}\(workload^2\) & 0.005 & \\
        \hspace{4pt} \textit{var\_u} & 0.110 & \\
        \hspace{4pt} \textit{var\_e} & 0.722 & \\
        \midrule
        Observations & 1268 & \\
        No. of course occurrences & 46 & \\
        
        \midrule
        \multicolumn{3}{l} {
            Standard errors in parentheses  
        } \\
        \multicolumn{3}{l} {
            \textbf{* p$<$0.05, ** p$<$0.01, *** p$<$0.001} 
        } \\
        
\bottomrule
\end{tabular}
\label{table:models}
\end{table*}

To determine how course type and control variables (student's motivation and perceived workload) relate to student's learning experience, all variables are considered jointly in a multilevel regression model (1). 

Table~\ref{table:models} presents the estimation results for the random coefficient model predicting student's learning experience. As shown, the coefficient for the course type B is negative and statistically significant (b $=$ -0.353, p $<$ 0.05) meaning that students rate their learning experience significantly worse in type B (software engineering process, models and methods) courses compared to type A courses (software construction and programming). Thus, Hypothesis 1 is supported and the results indicate that the type of course indeed has an effect on student evaluation of teaching in software engineering courses. When it comes to course type C (professional practices for software engineering), there is no statistically significant difference between the type C and type A courses in students' learning experience. Therefore Hypothesis 2 is only partly supported: Courses related to software construction and programming provide higher SET ratings than courses related to some, but not all, other knowledge areas. 

As to the control variables, student's motivation positively affects (b $=$ 0.525, p $<$ 0.001) learning experience. This finding is in line with previous observations that students’ motivation or subject interest is positively associated with SET~\cite{griffin2004grading, wachtel1998student}. In turn, the coefficient for the effect of squared workload on learning experience is negative (b $=$ -0.121, p $<$ 0.001), indicating that too light or too heavy workload leads to worse learning experience, and further confirming the findings of Centra~\cite{centra2003will} and Marsh~\cite{marsh2001distinguishing}.

\section{Discussion}

The objective of this study was to investigate the effect of the type (or topic) of a course on the student evaluations of teaching in a software engineering programme. We accomplished this goal by performing a multilevel modeling analysis on a set of 1295 evaluations collected from SE students from a total of 46 different course implementations. Previous research maintains that students value certain course topics, namely programming and software construction, over others. At the same time, universities often use SETs as their primary (or sole) metric for teaching quality. Therefore, establishing how the type (or topic) of a course affects the SETs is an important research topic.

In summary, our results suggest that the course type plays an essential role in SET responses of software engineering courses. It seems that students give higher ratings to software construction and programming related courses compared to some other knowledge areas when evaluating their learning experience. This result is in line with the works on student perceptions and expectations of the different SE topics.  

We used the SWEBOK knowledge areas as the basis of our course categorization. Regarding our hypothesis, however, there was no significant statistical difference between courses on the professional practices for software engineering (category C) and programming courses. This finding was somewhat surprising, since based on the established literature we expected that programming courses always get more positive ratings. This suggests that while the course type does have an affect on the SET ratings, there must also be other contributing factors that have a significant effect on the SETs. During preliminary analyses we tested the effect of two other level-2 factors, teaching language and course size, on learning experience. However, these variables were excluded from the final model due to collinearity with course type.

In the following sections, we first discuss implications of our findings and then address threats to validity. 

\subsection{Implications}
Our findings have implications for both the field of research in student perceptions of software engineering education, and the software engineering education practise.

First, our findings connect the field of student evaluation of teaching to the existing line of research in software engineering education that examined bias in what students consider to be essential courses~\cite{ivins2006software, hewner_undergraduate_2013, hewner_how_2014, gold2019software}.

Second, the findings have results for the practise of software engineering education, due to the fact that many universities use quantitative results from student surveys for their quality control and lecturer job performance processes. Our findings indicate that within our geographically limited dataset, students have bias towards programming courses and systematically give higher ``learning experience'' rating to practical programming courses. This bias cannot be attributed to lecturer as predictor, since examining course descriptions showed that lecturers taught multiple types of courses and sometimes switched courses over the duration of the datasets. Based on this, we publish a series of recommendations for practitioners who use SET in software engineering education:
\begin{itemize}
  \item \textit{Consider the effect of student bias when evaluating lecturers.} When using quantitative student evaluation of teaching to evaluate teachers, bias \cite{kohn2006role,feistauer2018validity} should be corrected against and current critique of SET methods should be considered before use.
  \item \textit{Evaluate the course evaluation instruments, and consider also utilizing qualitative metrics in addition to numeric data.} As established in SET research, SET metrics are seldom objective. Other, perhaps more qualitative metrics to evaluate teaching quality, should also be included in a holistic quality process. 
  \item \textit{Give students a comprehensive vision of software engineering work.} As part of introduction courses, software engineering students should be better introduced to the entire field and any misconceptions addressed.
  \item \textit{Connect software engineering theory to practise.} Fairly or not, students currently indicate (in the studied organization) that their ``learning experience'' was lower in theory-based courses. This might be due to the fact that in the studied organisation, most courses concentrate on a single topic. Can best practises from the software engineering education research community, such as problem-based learning \cite{ouhbi2020software}, be applied to connect theory with practise better?
\end{itemize}

The main limitation in this study is that the dataset is limited geographically to one organization and the findings cannot be generalized quantitatively. However, Urquhart \cite{urquhart2010putting} synthesizes a line of thinking and presents a concept of theoretical generalization \footnote{Also known as analytical generalization \cite{wohlin2012experimentation}}, where several qualitative or theory-based contributions are related to each other. From this perspective, our findings have wider utility in supporting similar individual findings from Ivins et al.~\cite{ivins2006software}, Hewner~\cite{hewner_undergraduate_2013}, and Gold-Veerkamp~\cite{gold2018using}. What is still required for future research is confirming that the student bias, shown to exist, affects student evaluation of teachers in other organizations.

\subsection{Threats to validity}
In this Section, we categorize and address threats to validity, following recommendations Wohlin et al.~\cite{wohlin2012experimentation} have summarized from the seminal work by Campbell et al.~\cite{nl1963experimental} and Yin~\cite{yin2017case}.
\paragraph*{Conclusion validity} The used statistical analysis methods are included in the best practises of survey outcome analysis~\cite{wohlin2012experimentation}. The analysis outcomes have sufficient statistical significance.
\paragraph*{Construct validity} The student feedback questionnaires used at the organization are based on accepted SET literature and constructs such as learning experience~\cite{clayson2009student}, motivation~\cite{griffin2004grading,marsh2009exploratory} and perceived workload~\cite{centra2003will}. Additionally, the course type categorization is based on SWEBOK and ACM curricula recommendations.
\paragraph*{Internal validity} The survey process itself is guided by the organization's quality assurance department and independent from the course lecturers. Student motivation and workload were controlled factors. While lecturer demographics or teaching methods were not controlled through the model, the department did not have a large quantity of the lecturers at the time and many lecturers taught courses across the SWEBOK categories. Furthermore, lecturers swapped courses during the data collection period, increasing diversity.
\paragraph*{External validity} The findings have been related to other studies in the field and confirm their findings using the principles of theoretical and analytical generalization.
\paragraph*{Reliability} The data analysis process was cross-checked by a team of three researchers with experience in the field. The data has been collected and validated by an independent quality assurance department.

\section{Conclusion}
To answer our research question, \textit{how does the type of a course affect student evaluation of teaching in software engineering courses}: The type (or topic) of the course can predict a higher SET rating. Software construction and programming courses receive higher SET ratings compared to some other topics. However, programming courses do not always provide a better SET rating. SET is a complex, multidimensional concept, and its validity for evaluating teaching quality is debated in the education research literature.  

This paper establishes, to our knowledge, the first explicit steps towards understanding the dimensions of SET in the software engineering education context. Our study extends the state of the art by synthesizing the established knowledge on how students tend to value some knowledge areas over others, and showing evidence of this in practice by analysing SET data. The results should provide insights about the use of SET for both software engineering educators and faculty administrators. In an increasingly data-driven world, we call on the education community to acknowledge the limitations and biases that exist in the way teaching quality is most commonly measured. 

In this paper, we follow Garcia-Martinez's 2010 call~\cite{garcia2010evaluation} for more research in student evaluation of teaching in fields related to computing that has mostly gone unanswered. We extend the state of the art by connecting findings in SET to the previous research of student bias by Ivins et al., Hewner, and Gold-Veerkamp~\cite{ivins2006software, hewner_undergraduate_2013, hewner_how_2014, gold2019software}. In this manner, we extend the scope of investigation from evaluating teacher characteristics~\cite{kavalchuk2020empirical} to systematic evaluation of course characteristics.

The main limitation of this paper is the geographically limited scope of the dataset. While the study covers data from multiple years, collecting data from other organizations would support generalizing the findings. For future research, we recommend wider replication studies to investigate if the phenomenon can be replicated in other software engineering programs and closely related fields.

\FloatBarrier

\bibliographystyle{IEEEtran}
\bibliography{references}

\end{document}